\documentclass[aps,pra,twocolumn,showpacs]{revtex4}
\usepackage{mathrsfs}
\usepackage{amssymb}
\usepackage{amsmath}
\usepackage{graphicx}
\usepackage{dcolumn}
\usepackage{bm}
\begin{document}

\title{Quantum geometric tensor and the topological characterization of the extended Su-Schrieffer-Heeger model}

\author{Xiang-Long Zeng}
\affiliation{School of Science, Beijing Information Science and Technology University, Beijing 100192, China;}

\author{Wen-Xi Lai}
\affiliation{School of Science, Beijing Information Science and Technology University, Beijing 100192, China;}

\author{Yi-Wen Wei}
\affiliation{School of Science, Beijing Information Science and Technology University, Beijing 100192, China;}

\author{Yu-Quan Ma*}
\affiliation{School of Science, Beijing Information Science and Technology University, Beijing 100192, China;}

\begin{abstract}
We investigate the quantum metric and topological Euler number in a cyclically modulated Su-Schrieffer-Heeger (SSH) model with long-range hopping terms. By computing the quantum geometry tensor, we derive exactly expressions for the quantum metric and Berry curvature of the energy band electrons, and we obtain the phase diagram of the model marked by the first Chern number. Furthermore, we also obtain the topological Euler number of the energy band based on the Gauss-Bonnet theorem on the topological characterization of the closed Bloch states manifold in the first Brillouin zone. However, some regions where the Berry curvature is identically zero in the first Brillouin zone results in the degeneracy of the quantum metric, which leads to ill-defined non-integer topological Euler numbers. Nevertheless, the non-integer "Euler number" provides valuable insights and provide an upper bound for absolute values of the Chern numbers.

\end{abstract}


\maketitle

\section{Introduction}

The Su-Schrieffer-Heeger (SSH) model is a topological quantum system model with a simple structure, but it has very typical topological properties \cite{1,2,3,4,5,6}, such as the winding number that characterizes the topological properties, the correspondence between bulk states and edge states, etc. \cite{7,8,9,10,11}. In addition, SSH models can be used to describe the one-dimensional polyacetylene, graphene ribbons \cite{12}, p-orbital light ladder systems \cite{13}, and off-diagonal two-color optical lattices \cite{14}. Historically, the Haldane model introduced a next-nearest neighbor inter-action in a two-dimensional honeycomb structure to realize the anomalous quantum Hall effect \cite{9}, causing the system to undergo a topological phase transition from an ordinary insulator to a Chern insulator. Interestingly, if we expand the SSH model by adding appropriate cyclic modulation parameters, we can obtain a phase diagram similar to the two-dimensional Haldane model \cite{15}, which further enriches the theoretical value of the one-dimensional SSH model and can be used to simulate two-dimensional topological systems \cite{16,17,18,19,20}. For the modulated SSH model, all parameters can be obtained by existing cold atom experimental techniques, optical systems or waveguide systems \cite{21,22,23,24}, such as the interaction of fermion atoms on the two-legged ladder \cite{13}, etc., then our results can also be verified by existing experiments.

As a general covariant tensor in Hilbert space geometry, the QGT ${Q_{\mu \nu }} = {g_{\mu \nu }} - \frac{i}{2}{F_{\mu \nu }}$ \cite{25,26} defined on a parameterized quantum state manifold is expected to shed some light on understanding quantum phase transitions in many-body systems \cite{27,28,29}. Its imaginary part (up to a coefficient)   is right the Berry curvature, which is a key quantity to derive the first Chern number in understand the topological quantum matter. Especially, the quantum metric  (real part of the QGT) proposed by Provost and Valee \cite{25} is a positive semi-definite Riemannian metric, which defines a gauge invariant distance between two adjacent quantum states in a parameterized Hilbert space. Recently, it has been shown that the quantum metric plays crucial roles in quantum transport phenomena, quantum noise, optical conductivity, anomalous Hall effect, unconventional superconductivity, and adjacent topics \cite{30,31,32,33,34,35,36,37,38,39,40,41,42,43,44,45,46,47,48,49,50,51,52,53,54,55,56,57}. Furthermore, it has been revealed that the quantum metric can provide a topological Euler number for the energy band, which is based on the Gauss-Bonnet theorem on the topological characterization of the closed Bloch states manifold in the first Brillouin zone. which provides an effective topological index for a class of nontrivial topological phases \cite{58,59,60,61,62,63,64,65,66,67,68,69,70,71,72}.

This paper is structured as follows. In Sec. 2, we study a cyclically modulated SSH model with long-range hop-ping terms, and solve its Hamiltonian in the Bloch momentum space with periodic conditions. In Sec. 3, we obtain the quantum geometric tensor for the occupied lower band of the extended SSH model. The critical points in the model can be witnessed by the singularity behaviors both of the Berry curvature and quantum metric. In Sec. 4, we study the topological Euler number of this model and make a comparison between the phase diagram marked by the Chern number and the Euler number, respectively. Finally, we provide a summary of our work. 

\section{The model}

We consider the SSH model with N lattice points, the Hamiltonian can be written as \cite{6}
\begin{equation}
\begin{split}
		H = \sum\limits_{j = 1}^N {({t_1}\hat a_j^\dag {{\hat b}_j} + {t_2}\hat b_j^\dag {{\hat a}_{j + 1}} + {t_3}\hat a_j^\dag {{\hat b}_{j + 1}}}\\
			+ {t_A}\hat a_j^\dag {\hat a_{j + 1}} + {t_B}\hat b_j^\dag {\hat b_{j + 1}} + H.c),
\end{split}
\end{equation}
where $\hat a_i^\dag ({\hat a_i})$ and $\hat b_i^\dag ({\hat b_i})$ are the creation (annihilation) operators at sublattice A and site B for each $j$-th cell, ${t_1}$, ${t_2}$ and ${t_3}$ represent the nearest-neighbor (NN), next-nearest-neighbor (NNN), and the third-nearest-neighbor (TNN) hopping amplitudes (see Figure 1).
\begin{figure}[h]
	\centering
	\includegraphics[scale=0.5]{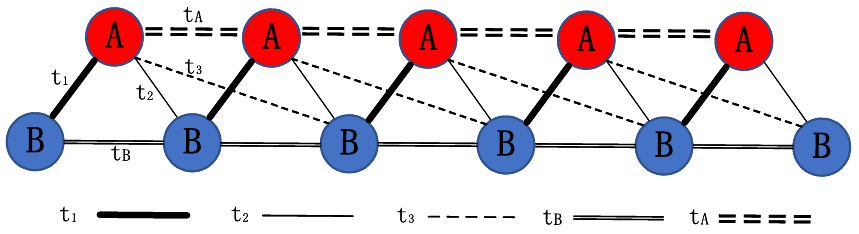}
	\caption{the extended SSH model of long-range interactions.}
\end{figure}

Now, we modulate ${t_1}$ and ${t_2}$
\begin{equation}
	\begin{split}
	{t_1} = 1 + \frac{1}{2}\cos \theta,\\
	{t_2} = 1 + \frac{1}{2}\cos \theta,
	\end{split}
\end{equation}
with ${\theta}$ representing a cyclical parameter varying from 0 to $2\pi$. Meanwhile, ${t_A}$ and ${t_B}$ are modulated
\begin{equation}
	\begin{split}
		{t_A} = h + \cos (\theta  + \phi ),\\
		{t_B} = h + \cos (\theta  - \phi ),
	\end{split}
\end{equation}
where $\phi $ is an additional parameter used to adjust the strength relation between ${t_A}$ and ${t_B}$. Then we consider the periodic boundary conditions and introduce the following Fourier transformation
\begin{equation}
	\begin{split}
	{\hat a_k} = \frac{1}{{\sqrt N }}\sum\limits_j {{\rm{ }}{e^{ikj}}{\rm{ }}{{\hat a}_j}},\\
	{\hat b_k} = \frac{1}{{\sqrt N }}\sum\limits_j {{\rm{ }}{e^{ikj}}{\rm{ }}{{\hat b}_j}},
	\end{split}
\end{equation}
and the Hamiltonian in momentum space can be written as
\begin{equation}
	H = \sum\limits_k {(\hat a_k^\dag ,\hat b_k^\dag ){\rm{ }}h(k,\theta ){\rm{ }}\left( {\begin{array}{*{20}{c}}
				{{{\hat a}_k}}\\
				{{{\hat b}_k}}
		\end{array}} \right)},
\end{equation}
where
\begin{equation}
	h(k,\theta ) = \left( {\begin{array}{*{20}{c}}
			{2{t_A}\cos k}&{{t_1} + {t_2}{e^{ - ik}} + {t_3}{e^{ik}}}\\
			{{t_1} + {t_2}{e^{ik}} + {t_3}{e^{ - ik}}}&{2{t_B}\cos k}
	\end{array}} \right)
\end{equation}
$h(k,\theta )$ can be written in the form of Pauli matrices
\begin{equation}
	h(k,\theta ) = \varepsilon \left( {k,\theta } \right){\rm{ }}{I_{2 \times 2}} + \sum\limits_{\alpha  = 1}^3 {{d_\alpha }(k,\theta ) \cdot {\sigma _\alpha }},
\end{equation}
where $\varepsilon \left( {k,\theta } \right)$ is the eigenvalue of the Hamiltonian, ${I_{2 \times 2}}$ is 2x2 identity matrix, ${d_\alpha }(k,\theta )$ is the coefficient of the Pauli matrix, ${\sigma _\alpha }$ is a Pauli matrix, representing pseudo-spin degrees of freedom. The diagonalization of $h(k,\theta )$ is straightforward and the eigenvalues can be written as
\begin{equation}
	E(k,\theta ) = \varepsilon (k,\theta ) \pm \sqrt {\sum\limits_{\alpha  = 1}^3 {d_\alpha ^2(k,\theta )} },
\end{equation}
and the eigenvectors is
\begin{equation}
	u(k,\theta ) = \frac{1}{{\sqrt {2d(d \mp {d_3})} }}{\rm{ }}\left( {\begin{array}{*{20}{c}}
			{{d_1} - i{d_2}}\\
			{ \pm d - {d_3}}
	\end{array}} \right),
\end{equation}

Here we choose the following modulated extended SSH model of long-range interactions as an example be-cause it has richer topological properties and higher Chern number topological phases than the ordinary SSH model. In this model, the Hamiltonian of the Bloch state momentum space is
\begin{eqnarray}
	\varepsilon \left( {k,\theta } \right) &=& (2h + 2\cos \theta \cos \varphi )\cos k,
	\notag \\
	{d_1} &=& (1 + \frac{1}{2}\cos \theta ) + (1 - \frac{1}{2}\cos \theta  + {t_3})\cos k,
	\notag \\
	{d_2} &=& (1 - \frac{1}{2}\cos \theta  - {t_3})\sin k,
	\notag \\
	{d_3} &=& (2h - 2\sin \varphi \sin \theta )\cos k.
\end{eqnarray}
It is obvious that the model is a generalized topological system. When ${t_3} = {t_A} = {t_B} = 0,$ and ${t_1} \ne 0$, ${t_2} \ne 0$, or when ${t_1}({t_2}) = {t_A} = {t_B} = 0$ and ${t_2}({t_1}) \ne 0$, the system will degenerate into a one-dimensional SSH model, but when ${t_3} = 0$ and ${t_A}$, ${t_B} \ne 0$, the system will become an extended SSH model of next-nearest neighbor interactions. Obviously, for ${t_A} = {t_B} = 0$ and ${t_1}$, ${t_2}$, ${t_3} \ne 0$, the model is equal to the two-coupled SSH models.
\begin{figure*}[tb]
	\begin{center}
		\includegraphics[width=0.35\textwidth]{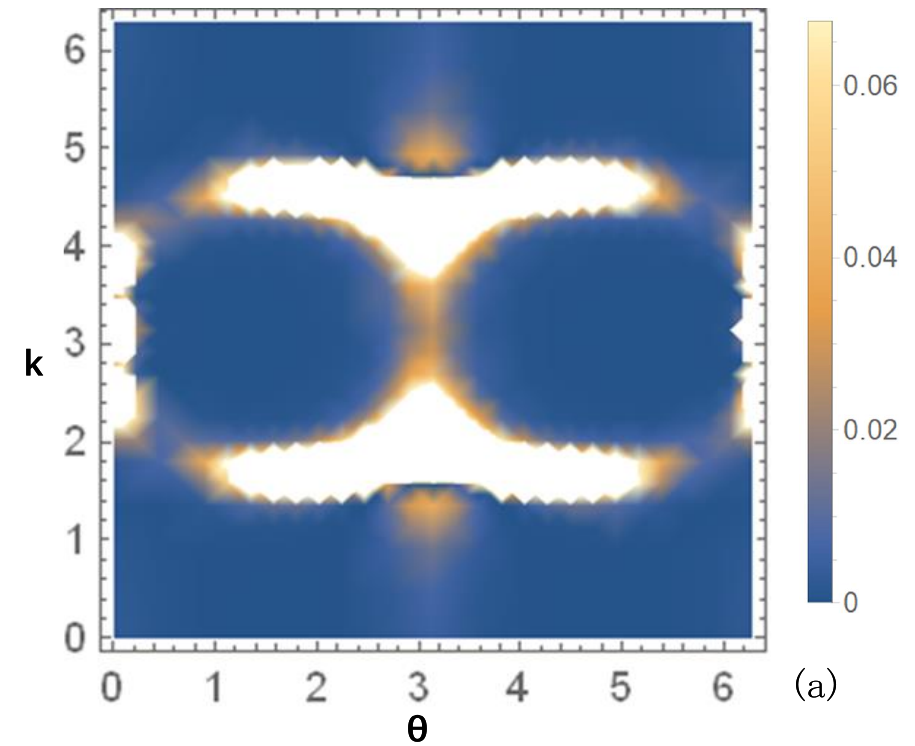} %
		\includegraphics[width=0.35\textwidth]{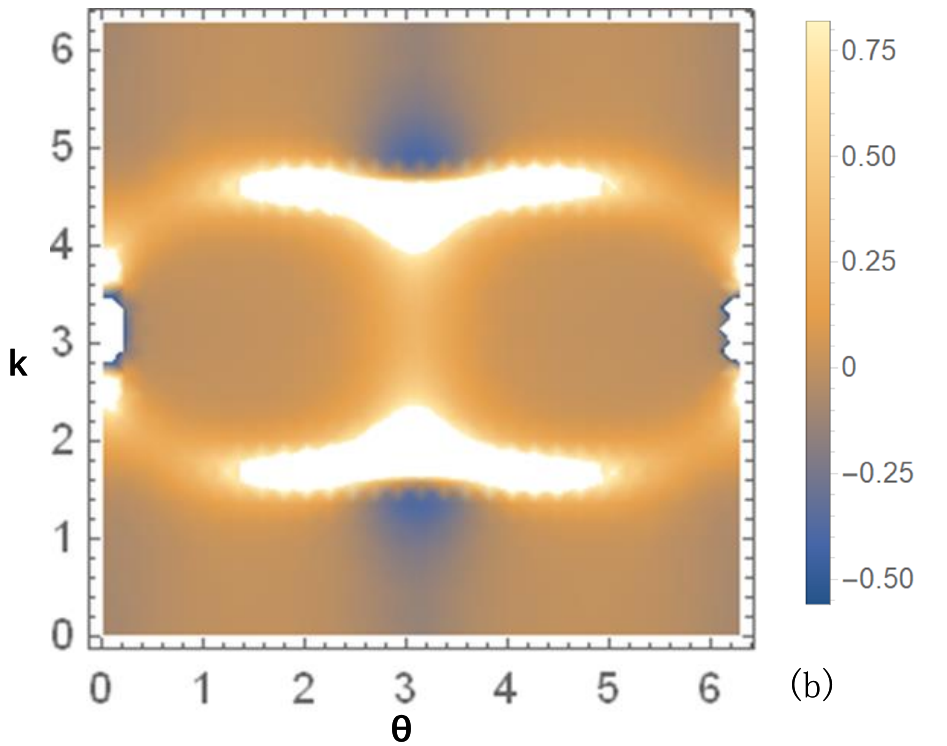} %
		\includegraphics[width=0.35\textwidth]{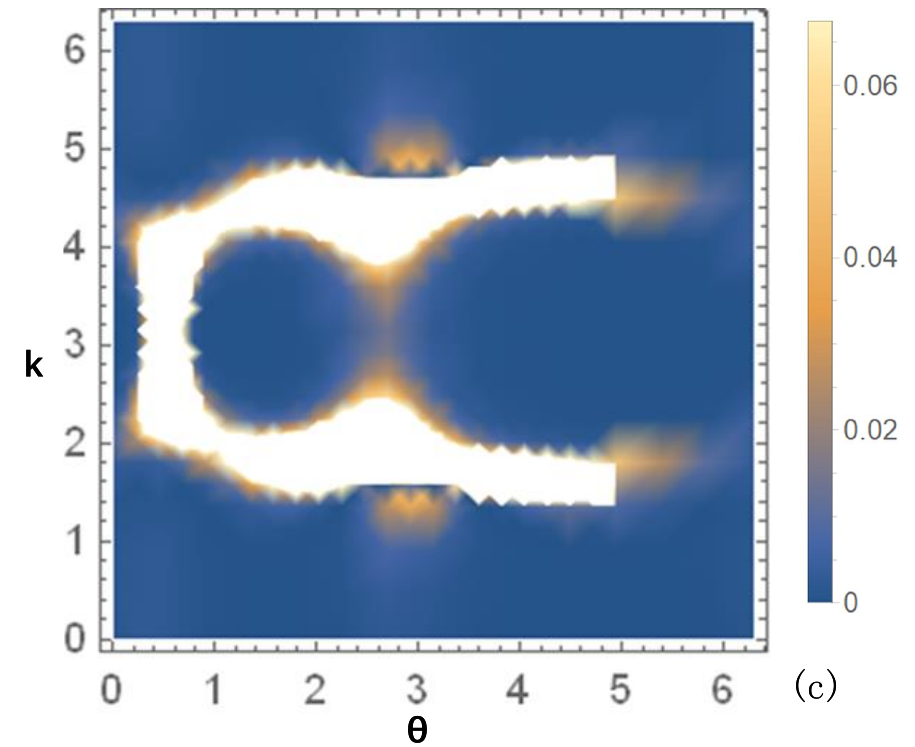} %
		\includegraphics[width=0.35\textwidth]{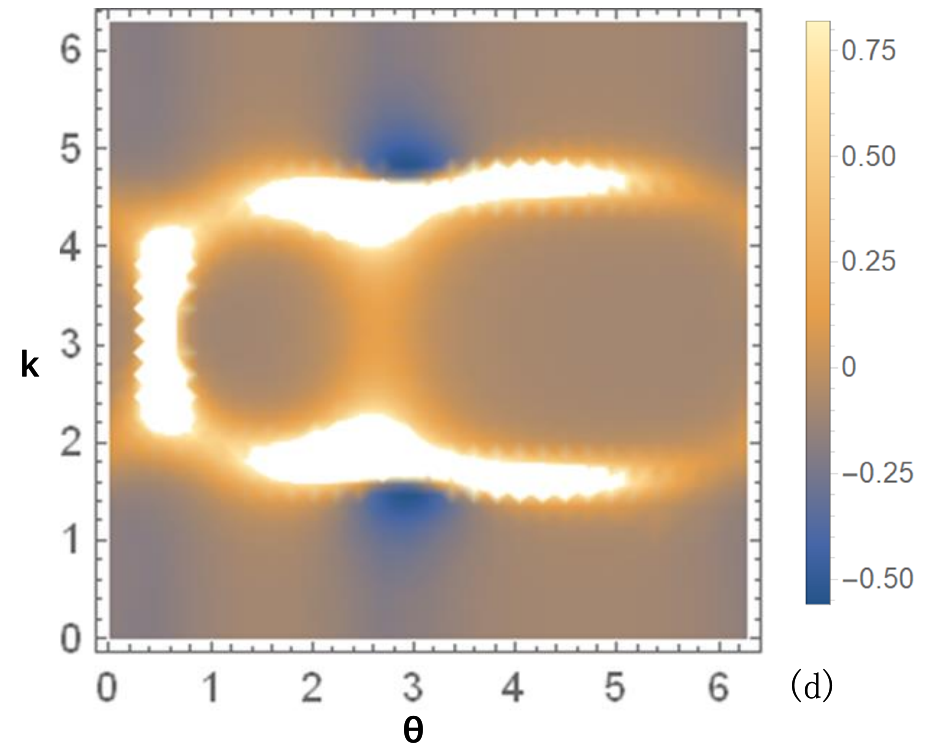} %
		\includegraphics[width=0.35\textwidth]{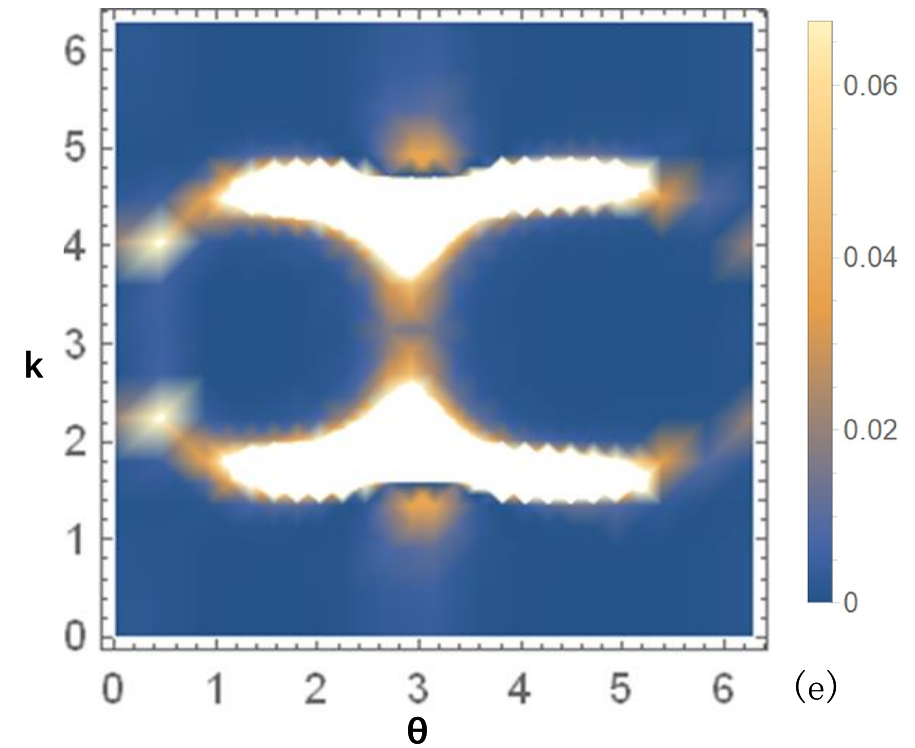} %
		\includegraphics[width=0.35\textwidth]{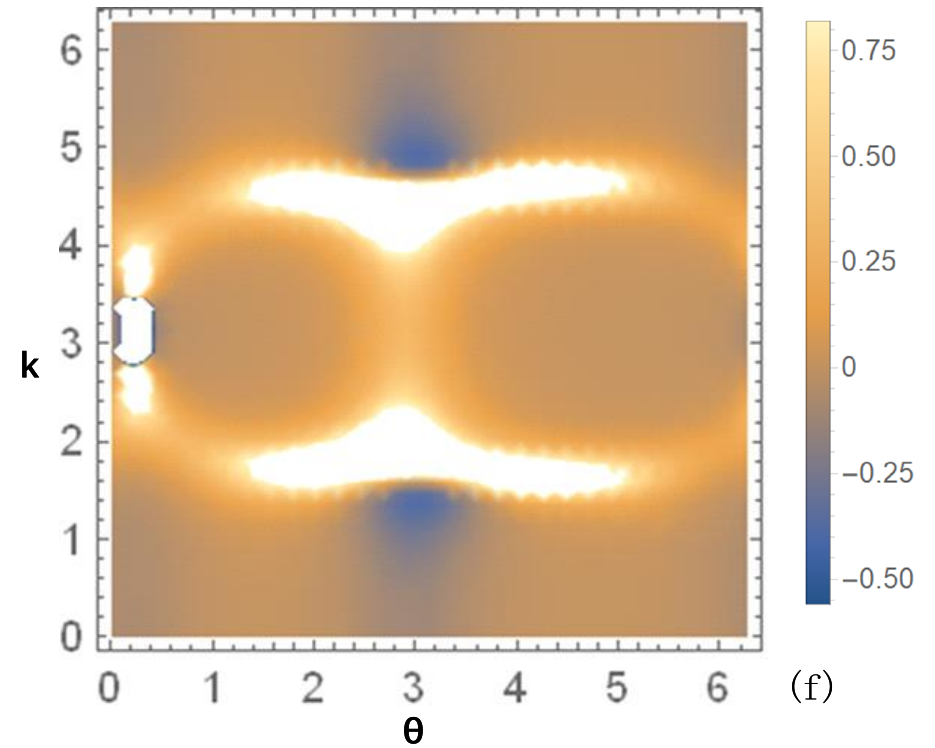}
	\end{center}
	\caption{The determinant of the quantum metric $\det {g_{\mu \nu }}$ and Berry curvature ${F_{\mu \nu }}$ as the functions of quasi-momentum $\theta ,k$ in the first Brillouin zone, with the different modulation parameters. (a) $\det {g_{\mu \nu }}$ with $h = 0.2$, $\phi  = \pi /2$, ${t_3} = 0.9$. (b)${F_{\mu \nu }}$ with $h = 0.2$, $\phi  = \pi /2$, ${t_3} = 0.9$. (c) $\det {g_{\mu \nu }}$ with $h = 1$, $\phi  = \pi /2$, ${t_3} = 1$. (d) ${F_{\mu \nu }}$ with $h = 1$, $\phi  = \pi /2$, ${t_3} = 1$. (e) $\det {g_{\mu \nu }}$ with $h = 0.5$, $\phi  = \pi /2$, ${t_3} = 0.9$. (f) ${F_{\mu \nu }}$ with $h = 0.5$, $\phi  = \pi /2$, ${t_3} = 0.9$.}
\end{figure*}

\begin{figure*}[tb]
	\begin{center}
		\includegraphics[width=0.35\textwidth]{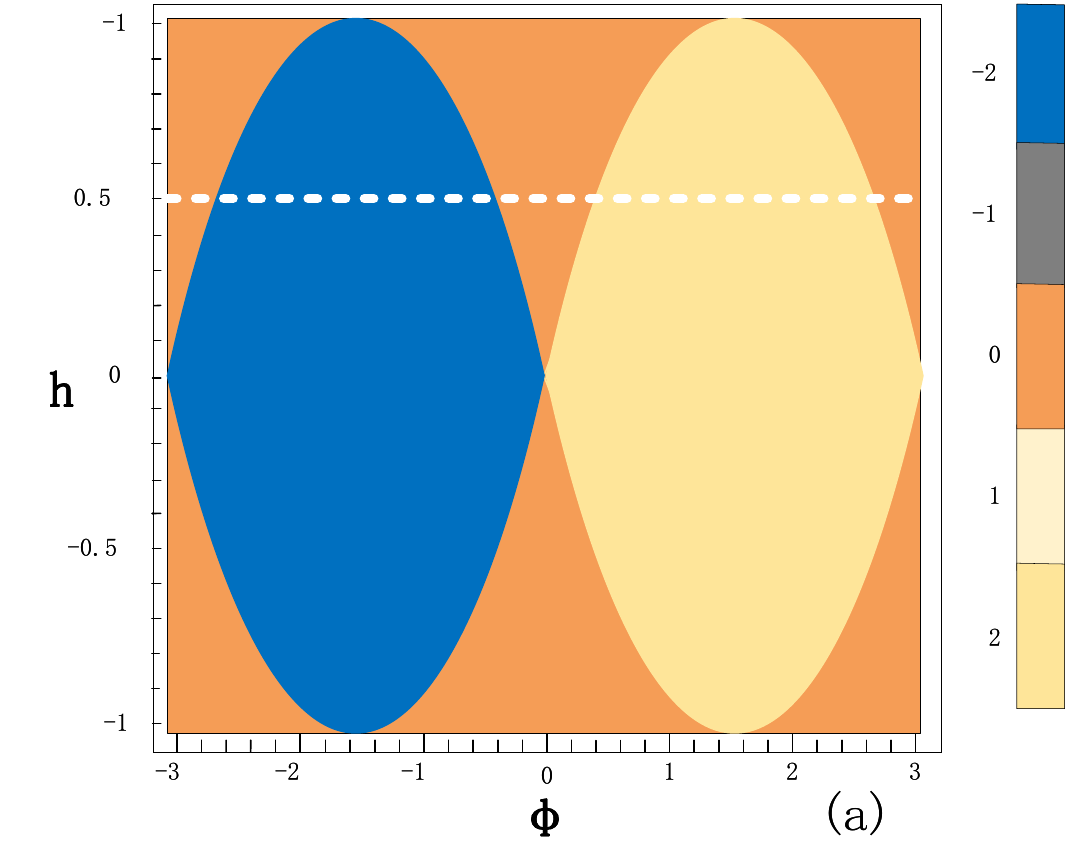} 
		\includegraphics[width=0.35\textwidth]{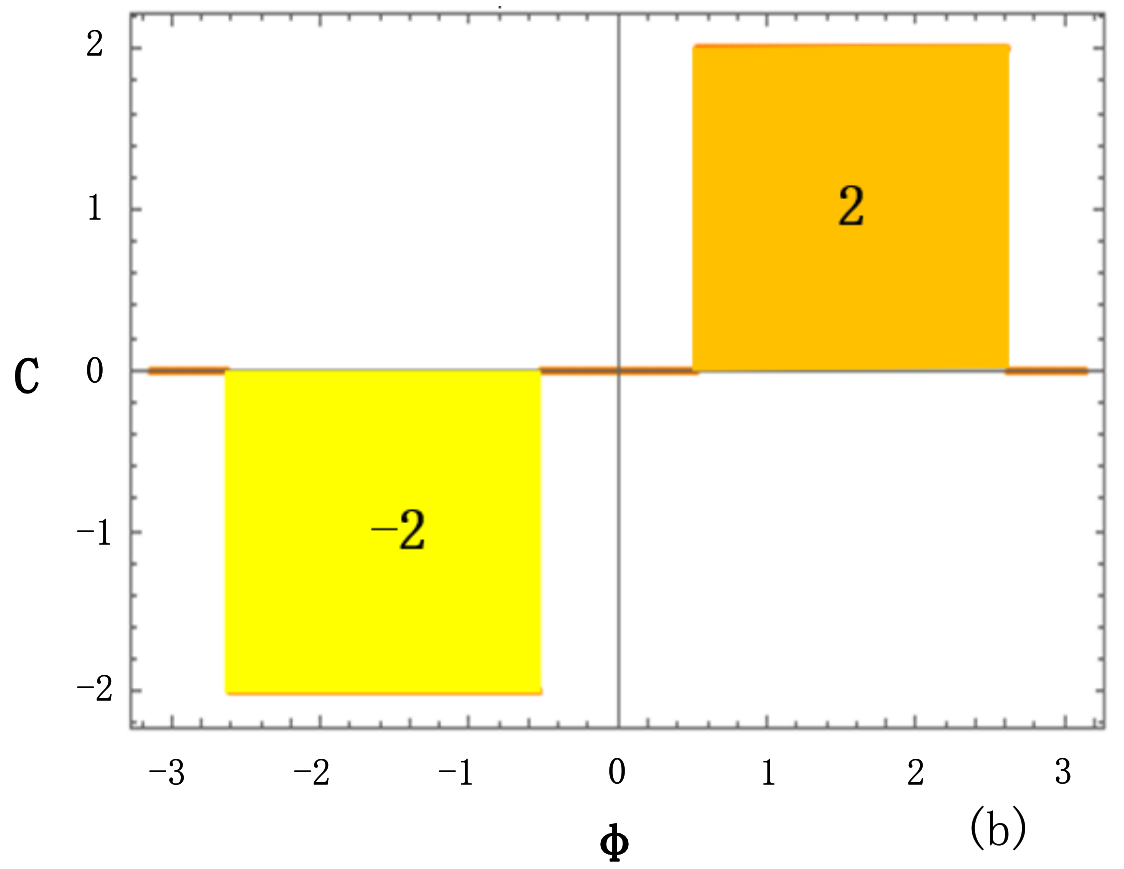} 
		\includegraphics[width=0.35\textwidth]{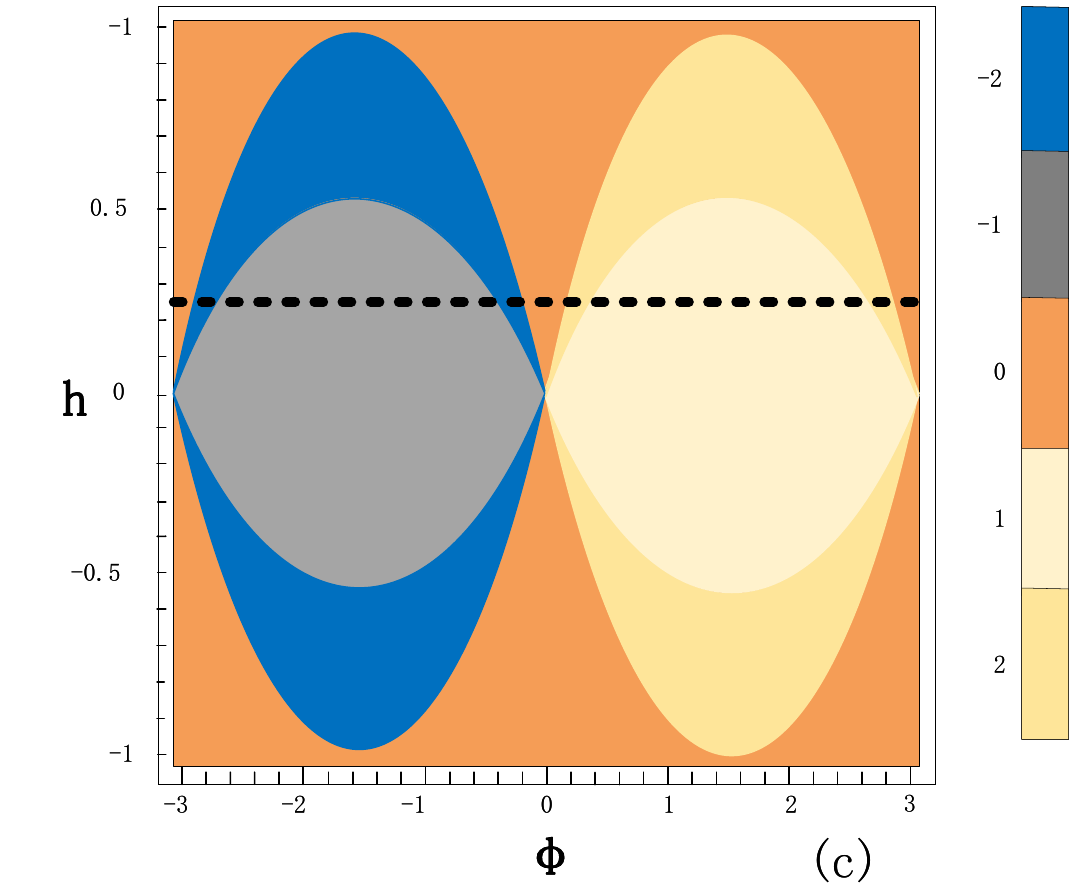} 
		\includegraphics[width=0.35\textwidth]{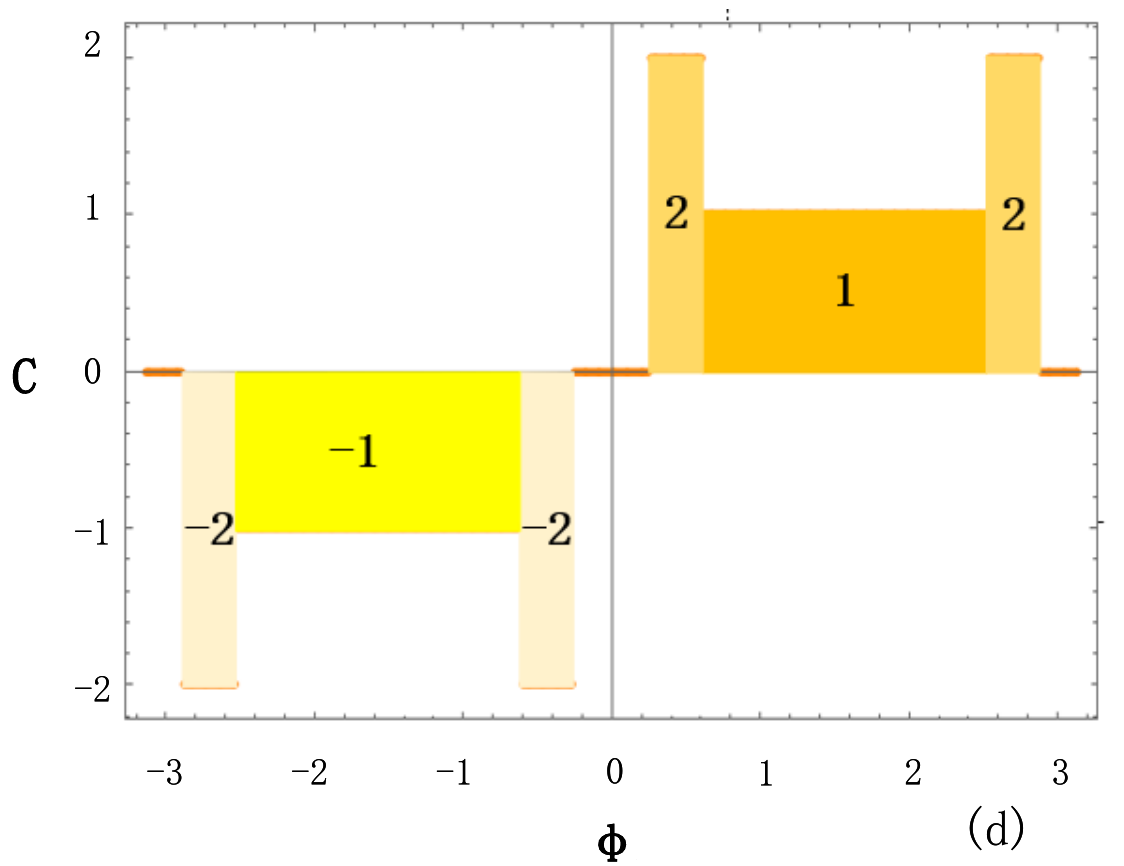} 
	\end{center}
	\caption{The Chern number varies with the parameters $h$ and $\phi$, and the other parameters are set to (a) ${t_3} = 1$. (b) The Chern number for the white line in Fig.3(a) where $h = \frac{1}{2}$ with $\phi\in(-3,3)$. (c) ${t_3} = 0.9$. (d) The Chern number for the white line in Fig.3(a) where $h = \frac{1}{4}$ with $\phi\in(-3,3)$.}
\end{figure*}
\section{Quantum metric, Berry curvature and the quantum geometric tensor}

Firstly, we introduce the quantum geometry tensor in the Bloch momentum space. It is derived from the gauge-invariant metric between two states on the $U(1)$ line bundle. We consider two close wave functions in the parameterized Hilbert space $\left| {\varphi (k)} \right\rangle $ and $\left| {\varphi (k + \delta k)} \right\rangle $, where $k = (\mu ,\nu )$ denotes the Hamiltonian parameters $(k, \theta)$ for convenience. The distance between two close wave functions is given by
\begin{equation}
	d{S^2} = \left\langle {{{\partial _\mu }\varphi (k){\rm{ }}d\mu }}
	\mathrel{\left | {\vphantom {{{\partial _\mu }\varphi (k){\rm{ }}d\mu } {{\partial _\nu }\varphi (k){\rm{ }}d\nu }}}
		\right. \kern-\nulldelimiterspace}
	{{{\partial _\nu }\varphi (k){\rm{ }}d\nu }} \right\rangle,
\end{equation}
the extending the factor $ \left| {{\partial _\mu }\varphi (k){\rm{ }}d\mu } \right\rangle $ to
\begin{equation}
	\left| {{\partial _\mu }\varphi (k){\rm{ }}d\mu } \right\rangle  = \left| {{D_\mu }\varphi (k)} \right\rangle  + [1 - P(k)]{\rm{ }}\left| {{\partial _\mu }\varphi (k)} \right\rangle,
\end{equation}
where $P(k) = \left| {\varphi (k)} \right\rangle \left\langle {\varphi (k)} \right|$ is projection operator, $\left| {{D_\mu }\varphi (k)} \right\rangle  = P(k){\rm{ }}\left| {{\partial _\mu }\varphi (k)} \right\rangle$ is the covariant derivative of $\left| {\varphi (k)} \right\rangle$. The quantum adiabatic approximation guarantees the parallel transport of the evolution of $\left| {\varphi (k)} \right\rangle$ to $\left| {\varphi (k + \delta k)} \right\rangle$ on the $U(1)$ line bundle, therefore $\left| {{D_\mu }\varphi (k)} \right\rangle  = 0$. We substitute Eq. (11) into Eq. (12) to obtain the quantum metric
\begin{equation}
	d{S^2} = \left\langle {{\partial _\mu }\varphi (k)} \right|[1 - P(k)]\left| {{\partial _\nu }\varphi (k)} \right\rangle d\mu d\nu, 
\end{equation}
And, the quantum geometry tensor is
\begin{equation}
	{Q_{\mu \nu }} = \left\langle {{\partial _\mu }\varphi (k)} \right|[1 - P(k)]\left| {{\partial _\nu }\varphi (k)} \right\rangle, 
\end{equation}
separating the quantum geometry tensor into the real and the imaginary parts, and we know that the real part is the quantum metric as ${g_{\mu \nu }} = {\mathop{\rm Re}\nolimits} {Q_{\mu \nu }}$,  and the imaginary part is 1/2 of the negative value of the Berry curvature as ${F_{\mu \nu }} =  - 2{\mathop{\rm Im}\nolimits} {Q_{\mu \nu }}$, thus we get the quantum geometry tensor for ${Q_{\mu \nu }} = {g_{\mu \nu }} - \frac{i}{2}{F_{\mu \nu }}$. And its imaginary part is canceled in the summation of the distance due to its antisymmetric, then the quantum metric can be rewritten as $d{S^2} = \sum\nolimits_{\mu \nu } {{\mathop{\rm Re}\nolimits} {Q_{\mu \nu }}d} \mu d\nu $, Given that $P(k) = \left| {\varphi (k)} \right\rangle \left\langle {\varphi (k)} \right|$, substitute it into Eq. (14) to get the quantum geometry tensor:
\begin{equation}
	{Q_{\mu \nu }} = \left\langle {{\partial _\mu }\varphi (k)} \right|{\rm{ }}[1 - \left| {\varphi (k)} \right\rangle \left\langle {\varphi (k)} \right|]{\rm{ }}\left| {{\partial _\nu }\varphi (k)} \right\rangle,
\end{equation}
substituting Eq. (15) into ${F_{\mu \nu }} =  - 2{\mathop{\rm Im}\nolimits} {Q_{\mu \nu }}$:
\begin{equation}
	{F_{\mu \nu }} = \left\langle {{{\partial _\mu }\varphi (k)}}
\mathrel{\left | {\vphantom {{{\partial _\mu }\varphi (k)} {{\partial _\nu }\varphi (k)}}}
	\right. \kern-\nulldelimiterspace}
{{{\partial _\nu }\varphi (k)}} \right\rangle  - \left\langle {{{\partial _\nu }\varphi (k)}}
\mathrel{\left | {\vphantom {{{\partial _\nu }\varphi (k)} {{\partial _\mu }\varphi (k)}}}
	\right. \kern-\nulldelimiterspace}
{{{\partial _\mu }\varphi (k)}} \right\rangle 
\end{equation}
then the Berry curvature can be calculated by Eq. (16) and Eq. (9):
\begin{equation}
	{F_{\mu \nu }} = \frac{1}{2}[{\bf{\hat d}} \cdot {\partial _\mu }{\bf{\hat d}} \times {\partial _\nu }{\bf{\hat d}}],
\end{equation}
where ${\bf{\hat d}}$ represents the unit vector ${\bf{d}}/d$ (see \cite{73} for details), and we calculate ${g_{\mu \nu }} = {\mathop{\rm Re}\nolimits} {Q_{\mu \nu }}$ according to Eq. (15)
\begin{eqnarray}
	{g_{\mu \nu }} &=& \frac{1}{2}\left\langle {{{\partial _\mu }\varphi (k)}}
	\mathrel{\left | {\vphantom {{{\partial _\mu }\varphi (k)} {{\partial _\nu }\varphi (k)}}}
		\right. \kern-\nulldelimiterspace}
	{{{\partial _\nu }\varphi (k)}} \right\rangle  + \frac{1}{2}\left\langle {{{\partial _\nu }\varphi (k)}}
	\mathrel{\left | {\vphantom {{{\partial _\nu }\varphi (k)} {{\partial _\mu }\varphi (k)}}}
		\right. \kern-\nulldelimiterspace}
	{{{\partial _\mu }\varphi (k)}} \right\rangle \notag \\
	&-& \left\langle {{{\partial _\mu }\varphi (k)}}
	\mathrel{\left | {\vphantom {{{\partial _\mu }\varphi (k)} {\varphi (k)}}}
		\right. \kern-\nulldelimiterspace}
	{{\varphi (k)}} \right\rangle \left\langle {{\varphi (k)}}
	\mathrel{\left | {\vphantom {{\varphi (k)} {{\partial _\nu }\varphi (k)}}}
		\right. \kern-\nulldelimiterspace}
	{{{\partial _\nu }\varphi (k)}} \right\rangle
\end{eqnarray}
The direct calculations of quantum metric ${g_{\mu \nu }}$ is tedious, however, it can be verified that there is a simple relation between the quantum metric determinant and the Bloch state $\left| {\varphi (k)} \right\rangle$ (for details see the Appendix A in Ref.\cite{57}):
\begin{equation}
	\sqrt {\det ({g_{\mu \nu }})}  = \sqrt {{{({\bf{\hat d}} \cdot {\partial _\mu }{\bf{\hat d}} \times {\partial _\nu }{\bf{\hat d}})}^2}/4},
\end{equation}
we can get the relationship between quantum metric and Berry curvature by comparing Eq. (17) and Eq. (19):
\begin{equation}
\det {g_{\mu \nu }} = \frac{1}{4}{({F_{\mu \nu }})^2}.
\end{equation}
In figure 2, we show the determinant of the quantum metric $\det {g_{\mu \nu }}$ and Berry curvature ${F_{\mu \nu }}$ as the functions of quasi-momentum $\theta ,k$ in the first Brillouin zone, with the different modulation parameters.

\section{Topological Euler number}
In the two-dimensional parameters $(\mu, \nu):=(k, \theta)$ space , the topology of the first Brillouin zone is a two-dimensional torus. Considering the Hamiltonian in the two-dimensional momentum space, the Bloch state $\left| {\varphi (k)} \right\rangle$ will adiabatically evolve a $U(1)$ line bundle. The first Chern number, which serves as a topological invariant for all filled bands, can be obtain by integrating the imaginary part (Berry curvature) of the quantum geometry tensor over the Brillouin zone.
\begin{equation}
	C = \frac{1}{{2\pi}}\sum\limits_n\iint_{BZ}{{F_{\mu \nu }}}{\rm{d}}\mu {\rm{d}}\nu.
\end{equation}
Here we assume that the model is half-filled and substitute Eq. (17) into Eq. (21)
\begin{equation}
	C = \frac{1}{{4\pi }}\iint{({\mathbf{\hat d}} \cdot {\partial _\mu}{\mathbf{\hat d}} \times {\partial _\nu }{\mathbf{\hat d}})}{\rm{d}}\mu {\rm{d}}\nu.
\end{equation}
As shown in Figure 3, the model can exhibit different topological phases with higher Chern numbers by varying the next-nearest-neighbor hopping term  , and undergoes the corresponding topological quantum phase transitions.

The topological Euler numbers can be derived from the Gauss-Bonnet theorem based on the quantum metric (real part of the quantum geometry tensor)
\begin{equation}
	\chi  = \frac{1}{{2\pi }}\int_{BZ} \mathcal{K} dA,
\end{equation}
where $\mathcal{K} = {R_{\mu \nu \mu \nu}}/\det g$ is the Gauss curvature, and $dA = {(\det g)^{1/2}}d\mu d\nu $ denotes the area measure according to the metric $g$, ${R_{\mu \nu \mu \nu}}$ is the covariant Riemannian curvature tensor. The direct calculation of the Gauss curvature is complicated, but it can be verified that there exists the following relation (for details see the Appendix B in Ref.\cite{57}) in a generalized two-band Hamiltonian on a 2D manifold as ${R_{\mu \nu \mu \nu}} = 4\det g$. Then, we generalize Eq. (23) to get the topological Euler number with energy band in the first Brillouin zone.
\begin{equation}
	\chi  = \frac{2}{\pi }\int_{BZ} {\sqrt {\det g} {\rm{ }}d\mu d} \nu.
\end{equation}
The numerical results of the topological Euler numbers have been shown in Figure 4 with the Hamiltonian of parameters ${t_3} = 1$, $h = \frac{1}{2}$.
\begin{figure}[h]
	\centering
	\includegraphics[scale=0.4]{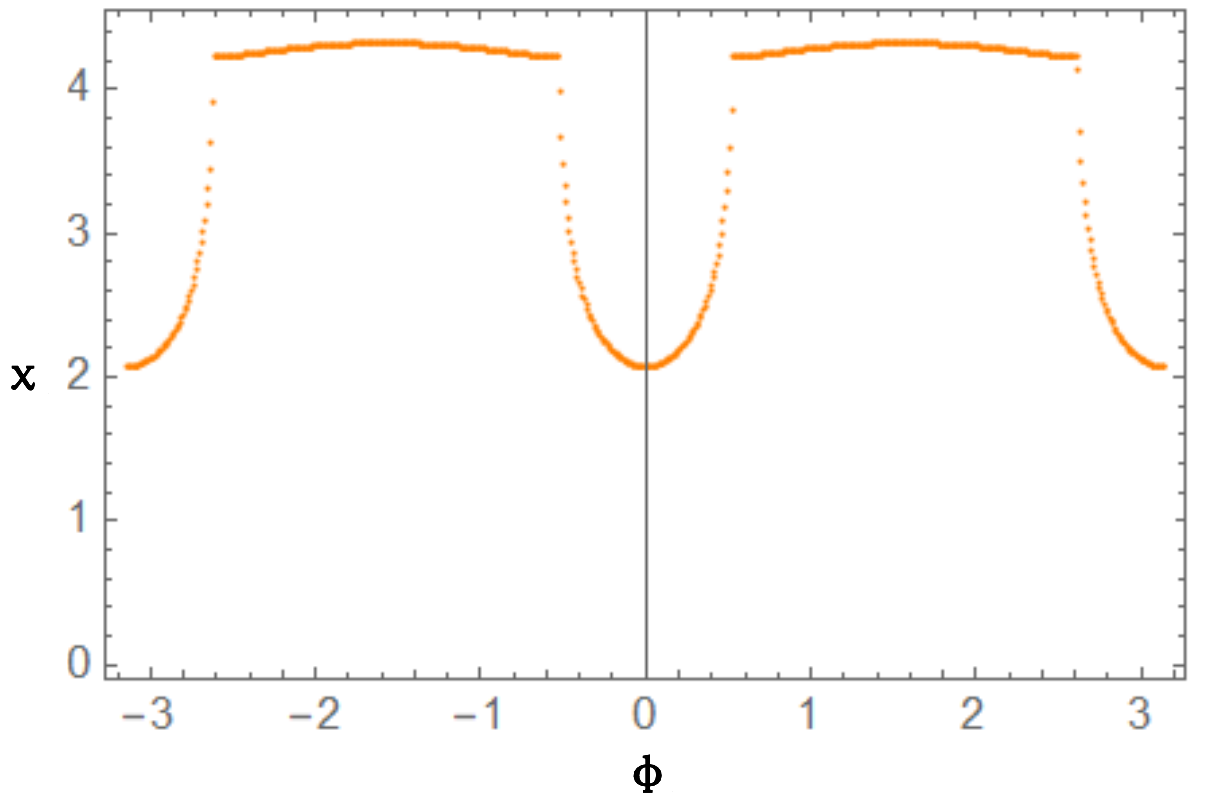}
	\caption{The Euler number varies with the parameter $\phi$, and the other parameters are set to ${t_3} = 1$, $h = \frac{1}{2}$.}
\end{figure}

As shown in Figure 4, the Euler number of the lower energy band is not exactly equal to 4 in the topological nontrivial phase where the first Chern number $=\pm 2$. The reason is that the quantum metric tensor is actually positive semi-definite. In a general two-dimensional two-band system, it can be proven that (for more details see Ref.\cite{Eulerint}): (1) If the phase is topological trivial, then the quantum metric must be degenerate~--- det$(g) =0 $ in some region of the first Brillouin zone. This leads to the invalidity of the Gauss-Bonnet formula and exhibits an ill-defined ``non-integer Euler number''; (2) If the phase is topological nontrivial with a non-vanishing Berry curvature, then the quantum metric will be a positive definite Riemann metric in the entire first Brillouin zone. Therefore the Euler number of the energy band will be guaranteed an even number $ 2(1-g) $ by the Gauss-Bonnet theorem on the closed two-dimensional Bloch energy band manifold with the genus $g$, which provides an effective topological index for a class of nontrivial topological phases.

In summary, we study the quantum geometry tensor and topological Euler number of an extended SSH model with long-range hopping terms. We show that the phase boundaries of the model can be witnessed by the singularity behaviors both of the Berry curvature and the quantum metric. We also study the topological Euler number of this model and make a comparison between the phase diagram marked by the Chern number and the Euler number, respectively. The degeneracy of the quantum metric in some regions of the first Brillouin zone leads to non-integer Euler numbers. However, the non-integer Euler number can also provide an upper bound for the corresponding Chern numbers \cite{Eulerint}.

\section{Acknowledgments}
Project supported by the Beijing Natural Science Foundation (Grant No. 1232026), the Qinxin Talents Program of BISTU (Grant No. QXTCP C201711), the R$\&$D Program of Beijing Municipal Education Commission (Grant No. KM202011232017), the National Natural Science Foundation of China (Grant No. 12304190), and the Research fund of BISTU (Grant No. 2022XJJ32).

\end{document}